\documentclass[aps,prx,reprint,superscriptaddress]{revtex4-1}
\usepackage{graphicx}
\usepackage{dcolumn}
\usepackage{bm}
\usepackage{natbib}
\usepackage{amsmath}
\usepackage{float}
\usepackage{hyperref}
\begin{document}

\preprint{}

\title{Spatially dispersing Yu-Shiba-Rusinov states in the unconventional superconductor $\mathrm{FeTe}_{0.55}\mathrm{Se}_{0.45}$}


\author{Damianos Chatzopoulos}
\thanks{These authors contributed equally to this work}
\affiliation{Leiden Institute of Physics, Leiden University, Niels Bohrweg 2, 2333 CA Leiden, The Netherlands}

\author{Doohee Cho}
\thanks{These authors contributed equally to this work}
\affiliation{Leiden Institute of Physics, Leiden University, Niels Bohrweg 2, 2333 CA Leiden, The Netherlands}
\affiliation{Department of Physics, Yonsei University, Seoul 03722, Republic of Korea}

\author{Koen M. Bastiaans}
\thanks{These authors contributed equally to this work}
\affiliation{Leiden Institute of Physics, Leiden University, Niels Bohrweg 2, 2333 CA Leiden, The Netherlands}
\thanks{These authors contributed equally to this work}

\author{Gorm O. Steffensen}
\affiliation{Center for Quantum Devices, Niels Bohr Institute, University of Copenhagen, Universitetsparken 5, 2100 Copenhagen Ø, Denmark.}

\author{Damian Bouwmeester}
\affiliation{Leiden Institute of Physics, Leiden University, Niels Bohrweg 2, 2333 CA Leiden, The Netherlands}
\affiliation{Kavli Institute of Nanoscience, Delft University of Technology, Lorentzweg 1, 2628 CJ Delft, Netherlands}

\author{Alireza Akbari}
\affiliation{Max Planck Institute for the Chemical Physics of Solids, D-01187 Dresden, Germany}
\affiliation{Max Planck POSTECH Center for Complex Phase Materials,and Department of Physics, POSTECH, Pohang, Gyeongbuk 790-784, Korea}

\author{Genda Gu}
\affiliation{Condensed Matter Physics and Materials Science Department, Brookhaven National Laboratory, Upton, NY 11973, USA}

\author{Jens Paaske}
\affiliation{Center for Quantum Devices, Niels Bohr Institute, University of Copenhagen, Universitetsparken 5, 2100 Copenhagen Ø, Denmark.}

\author{Brian M. Andersen}
\affiliation{Center for Quantum Devices, Niels Bohr Institute, University of Copenhagen, Universitetsparken 5, 2100 Copenhagen Ø, Denmark.}

\author{Milan P. Allan}
\email[]{allan@physics.leidenuniv.nl}
\affiliation{Leiden Institute of Physics, Leiden University, Niels Bohrweg 2, 2333 CA Leiden, The Netherlands}

\date{\today}

\begin{abstract}
By using scanning tunneling microscopy (STM) we find and characterize dispersive, energy-symmetric in-gap states in the iron-based superconductor $\mathrm{FeTe}_{0.55}\mathrm{Se}_{0.45}$, a material that exhibits signatures of topological superconductivity, and Majorana bound states at vortex cores or at impurity locations. We use a superconducting STM tip for enhanced energy resolution, which enables us to show that impurity states can be tuned through the Fermi level with varying tip-sample distance. We find that the impurity state is of the Yu-Shiba-Rusinov (YSR) type, and argue that the energy shift is caused by the low superfluid density in $\mathrm{FeTe}_{0.55}\mathrm{Se}_{0.45}$, which allows the electric field of the tip to slightly penetrate the sample. We model the newly introduced tip-gating scenario within the single-impurity Anderson model and find good agreement to the experimental data.
\end{abstract}



\maketitle

\section{Introduction}
The putative $s_{\pm}$ superconductor $\mathrm{FeTe}_{0.55}\mathrm{Se}_{0.45}$ is peculiar because it has a low Fermi energy and an unusually low and inhomogeneous superfluid density \cite{Bendele2010,Homes2015,Cho2019,Lubashevsky2012,Rinott2017,Miao2018}. It has been predicted to host a topological superfluid and Majorana zero-mode states \cite{Wu2016,Wang2015,Xu2016}. These predictions have been supported by recent experiments: photoemission has discovered Dirac-like dispersion of a surface state \cite{Zhang2018} while tunneling experiments have concentrated on in-gap states in vortex cores, which have been interpreted as Majorana bound states \cite{Wang2018,Zhu2020} since the low Fermi energy allows to distinguish them from conventional low-energy Caroli-Matricon-de Gennes states \cite{Chen2018}.

In-gap states have a long history of shining light into the properties of different host materials, and have allowed to bring insight into gap symmetry and structure, symmetry breaking, or the absence of scattering in topological defects, to name a few~\cite{Huang2019,Menard2015,Heinrich2018,Hudson2001,Zhou2013,Hoffman2002,Allan2013a,Allan2013,Rosenthal2014,Beidenkopf2011}. 
Impurity bound states have also been investigated in chains or arrays of magnetic impurities on superconducting surfaces where they can lead to Majorana edge-states \cite{Sticlet2019,Palacio-Morales2019,Poyhonen2018,Nadj-Perge2014}. In the case of $\mathrm{FeTe}_{0.55}\mathrm{Se}_{0.45}$, zero-bias in-gap resonances have become a primary way to identify Majorana bound states at magnetic impurity sites  or in vortex cores. At impurity sites, robust zero-bias peaks have been reported at interstitial iron locations which suggest the presence of Majorana physics \cite{Yin2015}. In addition, very recently STM experiments reported signatures of reversibility between magnetic impurity bound states and Majorana zero modes by varying the tip-sample distance on magnetic adatoms \cite{Fan2020}. Interestingly, there have also been signatures of spatially varying in-gap impurity states \cite{Machida2019,Wang2020} which are not yet understood.

Here we report the detection of in-gap states at sub-surface impurities, which are spatially dispersing, i.e.\ they change energy when moving away from the impurity site by a distance of $\Delta y$. The energy can also by tuned by changing the tip-sample distance ($\Delta d$). We argue that the most likely explanation of our observations involves a magnetic impurity state of the YSR type affected by the electric field of the tip. We show good agreement between our experimental findings and the single impurity Anderson model.

\begin{figure*}[ht]
\includegraphics[scale=1]{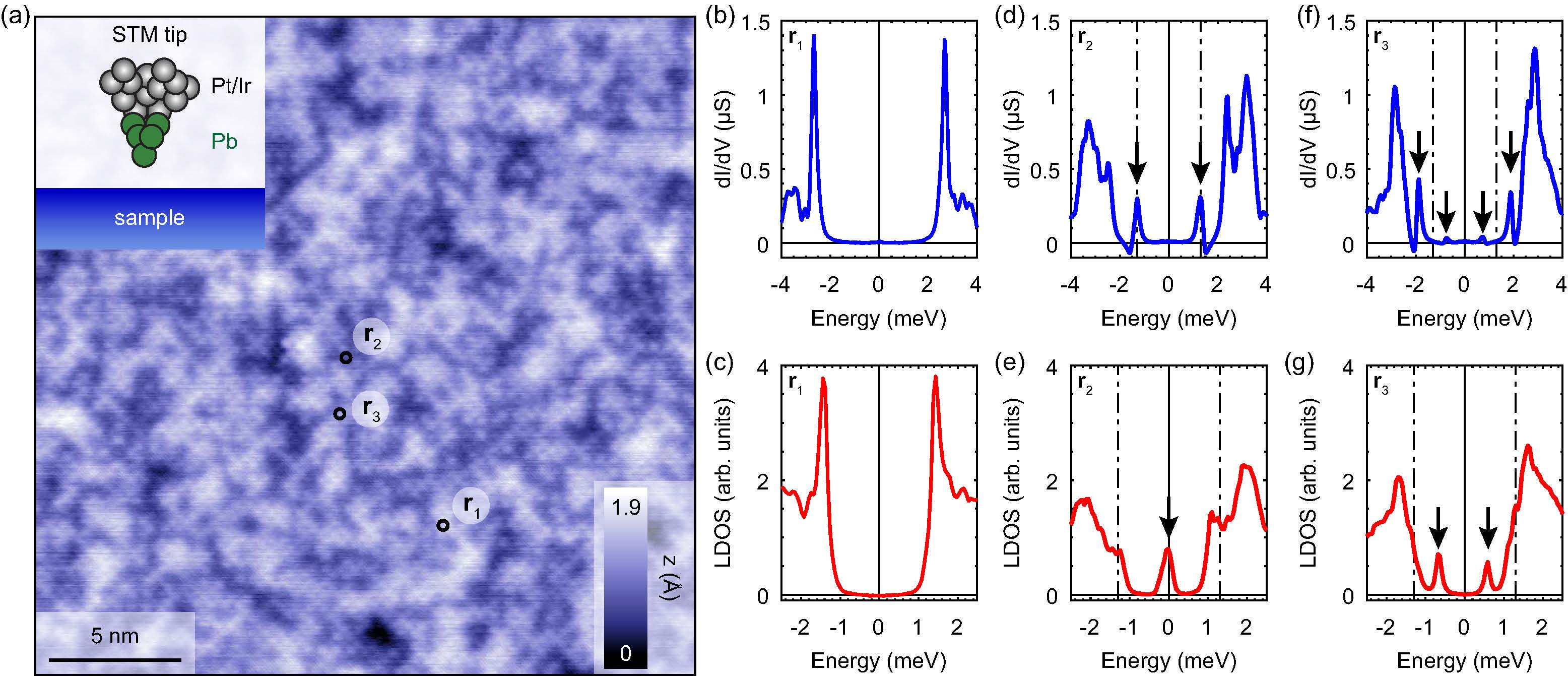}
\caption{Scanning Tunneling Microscopy on $\mathrm{FeTe}_{0.55}\mathrm{Se}_{0.45}$ with a superconducting tip. (a) Atomically resolved topographic image (25 $\times$ 25 nm$^{2}$) of $\mathrm{FeTe}_{0.55}\mathrm{Se}_{0.45}$ cleaved surface acquired with a Pb coated Pt/Ir tip (see inset) at 2.2 K in ultra-high vacuum. Setup condition: $V_{\mathrm{set}} = -8$ mV, $I_{\mathrm{set}} = -100 $ pA. (b),(d),(f) Average differential conductance spectra in the areas ($\mathbf{r}_{1}$, $\mathbf{r}_{2}$, $\mathbf{r}_{3}$) marked by the black circles in (a). $\mathbf{r}_{1}$: no in-gap states. $\mathbf{r}_{2}$: two in-gap resonances at $\pm 1.3$ meV. $\mathbf{r}_{3}$: two sets of symmetric peaks around the Fermi level. (c),(e),(g) Deconvolution of the  spectra shown in (b),(d),(f), respectively, provide information about the intrinsic LDOS of the sample in the indicated areas. In $\mathbf{r}_{2}$ a zero-bias impurity state is recovered and in $\mathbf{r}_{3}$ two in-gap states are observed. Setup conditions: (b) $V_{\mathrm{set}} = 6$ mV, $I_{\mathrm{set}} = 1.2$ nA, (d),(f) $V_{\mathrm{set}} = 5$ mV, $I_{\mathrm{set}} = 2$ nA. Lock-in modulation is $V_{\mathrm{mod}}= 30$ $\mu$V peak-to-peak for all measured spectra.}\label{fig1}
\end{figure*}

\section{Results and discussion}
\subsection{Detection of a particle-hole symmetric in-gap state in $\mathrm{FeTe}_{0.55}\mathrm{Se}_{0.45}$}
We use $\mathrm{FeTe}_{0.55}\mathrm{Se}_{0.45}$ samples with a critical temperature of $T_{\mathrm{C}}=14.5$ K. They are cleaved at $\sim 30$ K in ultra-high vacuum, and immediately inserted into a modified Unisoku STM at a base temperature of 2.2 K. To increase the energy resolution, we perform all tunneling experiments using a superconducting tip, made by indenting mechanically grinded Pt-Ir tips into a clean Pb(111) surface. With the superconducting tip and to leading order in the tunnel coupling, the current-voltage ($I-V$) characteristic curves are proportional to the convolution of the density of states of Bogoliubov quasiparticles in the tip and the sample
\begin{multline}
I(\mathbf{r},V)\sim \int D_{\mathrm{t}}(\omega+eV)D_{\mathrm{s}}(\mathbf{r},\omega)\times \\ [f(\omega,T)-f(\omega+eV,T)]d \omega,
\end{multline}
where $D_{\mathrm{s(t)}}$ is the density of states of the quasiparticles in the sample (tip), $f(\omega,T)$ is the Fermi-Dirac distribution at temperature $T$ and $e$ is the electron charge. In such a superconducting tunnel junction the coherence peaks in the conductance spectra, $dI/dV(\mathbf{r},V)$, appear at energies: $\pm (\Delta_{\mathrm{t}}+\Delta_{\mathrm{s}})$, where $\Delta_{\mathrm{s(t)}}$ is the quasiparticle excitation gap of the sample (tip). In addition, the energy resolution is far better than the conventional thermal broadening of $\sim 3.5 k_{\mathrm{B}}T$ ($k_{\mathrm{B}}$ is the Boltzmann constant) since it is enhanced by the sharpness of the coherence peaks of  $D_{\mathrm{t}}$ \cite{Franke2011}. To obtain the intrinsic local density of states (LDOS) of the sample, $D_{\mathrm{s}}(\mathbf{r},\omega)$, we numerically deconvolute our measured $dI/dV(\mathbf{r},V)$ spectra while retaining the enhanced energy resolution (for more details see Supplemental Material \cite{Supplement}). For this, we use our knowledge of the density of states of the tip with a gap of $\Delta_{\mathrm{t}}=1.3$ meV from test experiments on the Pb(111) surface using the same tip.

Figure \ref{fig1}(a) shows a topography of the cleaved surface of $\mathrm{FeTe}_{0.55}\mathrm{Se}_{0.45}$ obtained with a Pb coated tip (see inset). Brighter (darker) regions correspond to Te (Se) terminated areas of the cleaved surface which has a tetragonal crystal structure. Spatially resolved scanning tunneling spectroscopy shows that most locations have a flat gap as shown in Figs. \ref{fig1}(b)-(c). However, when we acquire spectra at specific points indicated by black circles ($\mathbf{r}_{2}$ and $\mathbf{r}_{3}$) in Fig. \ref{fig1}(a), we find sharp in-gap states. Figures \ref{fig1}(d)-(e) and \ref{fig1}(f)-(g) show such states, both in the raw data as well as in the deconvoluted results. The measured in-gap state is symmetric in energy, i.e. it is visible at $\pm E_{\mathrm{ig}}$ (Fig. \ref{fig1}(g)), or at the Fermi level, $E_{\mathrm{ig}} = 0$ (Fig. \ref{fig1}(e)). In the raw data (before numerical deconvolution) the states are located at energies $\pm(\Delta_{\mathrm{t}} \pm E_{\mathrm{ig}})$ (see arrows in Fig. \ref{fig1}(d) and \ref{fig1}(f)) due to the use of the superconducting tip. 

\begin{figure*}[ht]
\includegraphics[scale=1]{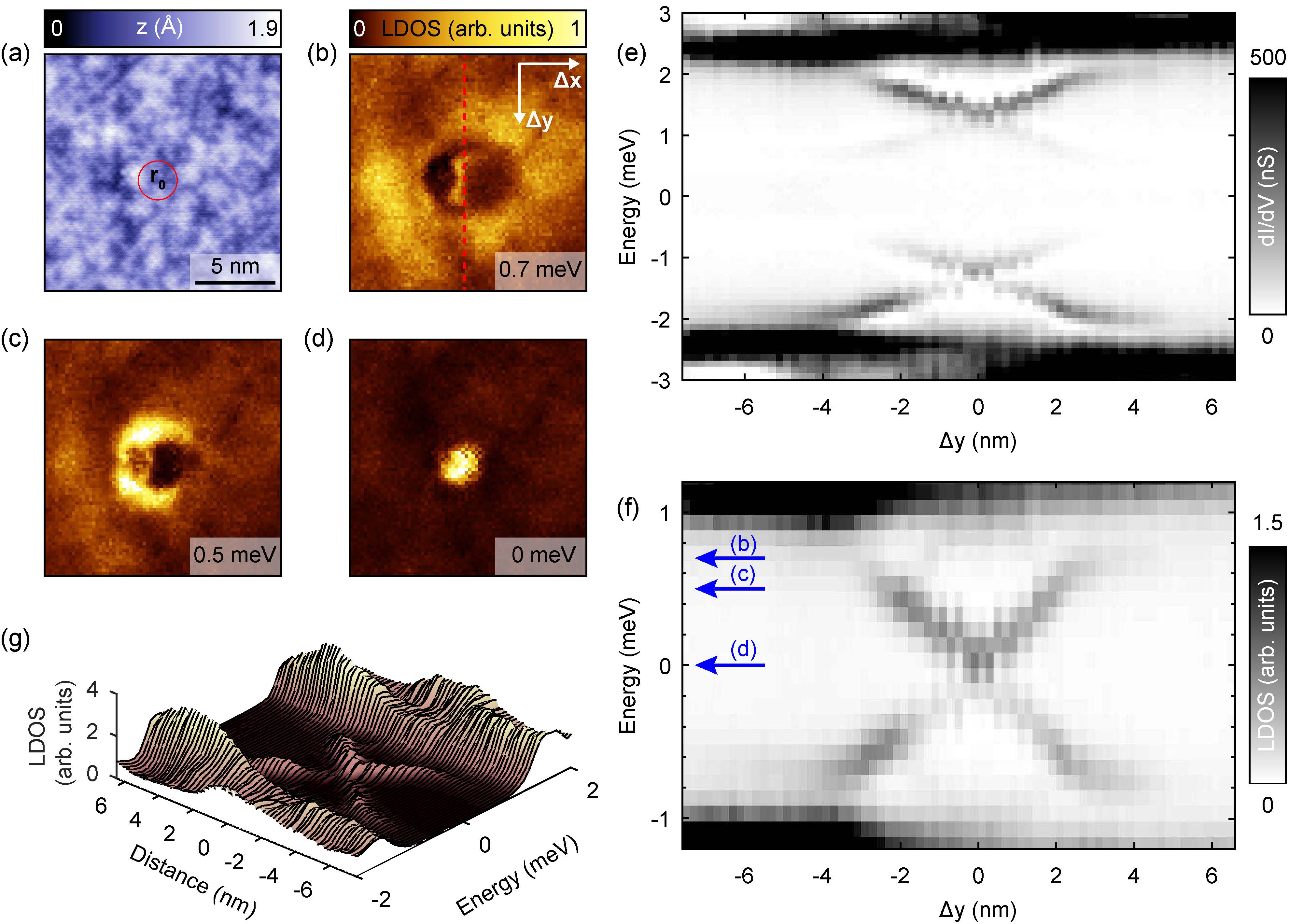}
\caption{X-shaped spatial dispersion of impurity resonances in $\mathrm{FeTe}_{0.55}\mathrm{Se}_{0.45}$. (a) Topographic image at the impurity location ($\mathbf{r}_{0}$ indicates the impurity center). No clear signature of the impurity is observed. Setup conditions: $V_{\mathrm{set}} = -8$ mV, $I_{\mathrm{set}} = -100$ pA. (b)-(d) Spatially resolved LDOS maps at different energies obtained by deconvolution of a $dI/dV(\mathbf{r},V)$ map in the same field-of-view as in (a). The energy of each LDOS map is indicated at the bottom right corner. (e) Measured differential conductance intensity plot of a vertical linecut passing through the impurity center $\mathbf{r}_{0}$ ($\Delta y = 0$ nm). The linecut was taken along the red dashed line in (b). A crossing of the in-gap resonances at the impurity center is observed. Setup conditions: $V_{\mathrm{set}} = -8$ mV, $I_{\mathrm{set}} = -1.6$ nA. Lock-in modulation is $V_{\mathrm{mod}}= 100$ $\mu$V peak-to-peak, (f) Deconvolution of the measured spectra in (e) shows an X-shaped dispersion of the sub-gap states crossing the Fermi level at the impurity center. The blue arrows indicate the energy of the maps in (b)-(d). (g) Series of LDOS spectra depicting the X-shaped spatial dispersion shown in (f).}\label{fig2}
\end{figure*}

\subsection{Spatial dispersion of the in-gap state}

In order to characterize the impurity in more detail we acquire a spatially resolved $dI/dV(\mathbf{r},V)$ map in the area shown in Fig. \ref{fig2}(a). Three energy layers of the deconvoluted map depicting the LDOS variations are shown in Figs. \ref{fig2}(b)-(d). The impurity exhibits a clear ring-shaped feature which eventually becomes a disk with smaller radius at the Fermi level. A spatial line cut profile along the red dashed line shown in Fig. \ref{fig2}(b) reveals two symmetric resonances around zero energy that extend over $\sim 10$ nm in space (Fig. \ref{fig2}(e)). Importantly, the energies of the in-gap states vary spatially as shown in the spatial cuts (Figs. \ref{fig2}(e)-(g)) obtained from the same conductance map. The dispersion of the in-gap states shows an X-shaped profile where the crossing point is indicated with $\mathbf{r}_{0}$ (Fig. \ref{fig2}(a)). In more detail, the state is at zero energy at $\mathbf{r}_{0}$, and then moves away from the Fermi level, before fading out slightly below the gap edge. We will show later that the character of this dispersion is dependent on the tip-sample distance, and that there can also be zero or two crossing points. By inspecting the topography at $\mathbf{r}_{0}$ we find no signature of irregularities, which points towards a sub-surface impurity defect as the cause of the observed in-gap peaks in the spectra. Similar observations have been reported previously on $\mathrm{FeTe}_{0.55}\mathrm{Se}_{0.45}$, but without a clear energy cross at the Fermi level \cite{Machida2019,Wang2020}. 

\subsection{YSR impurity states}
Our observations are reminiscent of YSR states caused by magnetic impurities in conventional superconductors \cite{Yu1965,Shiba1968,Rusinov1969,Franke2011,Ruby2015}. When a single magnetic impurity is coupled to a superconductor with energy gap $\Delta$ via an exchange coupling $J$ then the ground state of the many-body system depends on the interplay between superconductivity and the Kondo effect (described by the Kondo temperature $T_{\mathrm{K}}$). For $\Delta \gtrsim k_{\mathrm{B}}T_{\mathrm{K}}$ the superconducting ground state prevails (unscreened impurity) whereas for $\Delta < k_{\mathrm{B}}T_{\mathrm{K}}$ the Kondo ground state dominates (screened impurity). In each case, quasiparticle excitations above the ground state give rise to resonances symmetrically around the Fermi level inside the superconducting gap. In an STM experiment, this results in peaks in the conductance spectrum at the energy of the two YSR excitations which is determined by the product $\nu_{\mathrm{F}}JS$, where $S$ is the impurity spin and $\nu_{\mathrm{F}}$ the normal state density of states in the superconducting host ($\mathrm{FeTe}_{0.55}\mathrm{Se}_{0.45}$ in our case).

\begin{figure*}[ht!]
\includegraphics[scale=1]{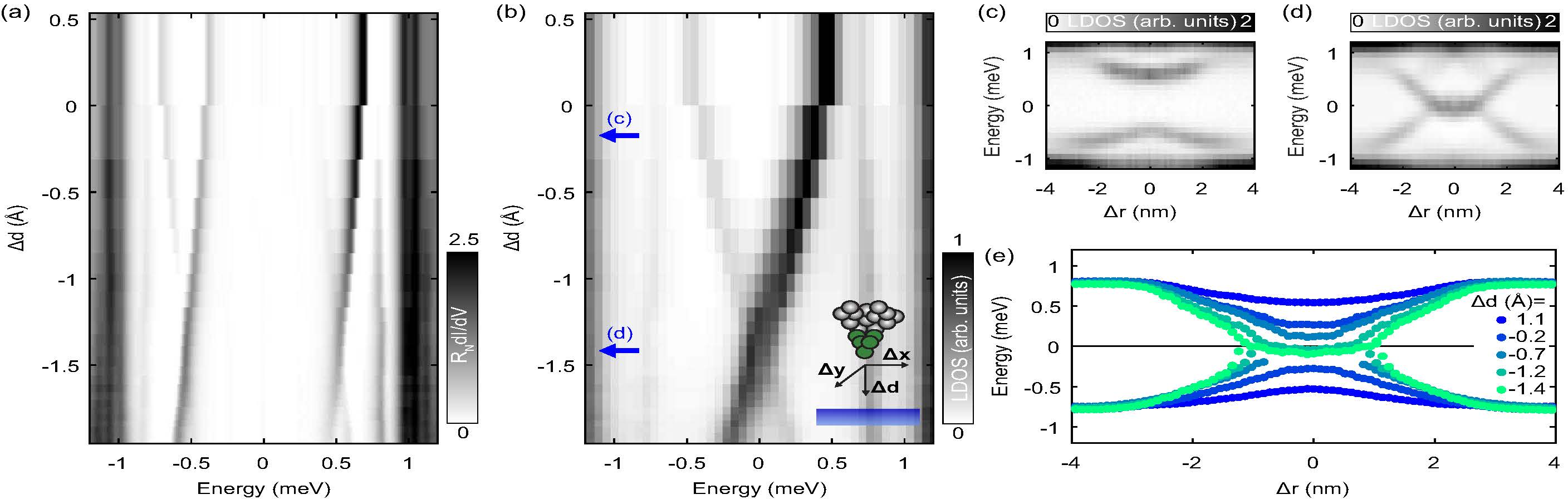}
\caption{Energy dispersion of the sub-gap resonances as a function of the tip-sample distance. (a) Conductance intensity plot for varying tip-sample distance ($\Delta d$) normalized by the normal state resistance $R_{\mathrm{N}}$. The in-gap states disperse and cross at the Fermi level. (b) Same as in (a) for the deconvoluted LDOS data. Inset: schematic representation of the tip movement when we vary the tip-sample distance ($\Delta d$) and when the tip scans laterally ($\Delta y$ and $\Delta x$) at a constant height. (c)-(d) Azimuthally-averaged radial profiles at different tip-sample distances indicated by blue arrows in (b). $\Delta r$, indicates the radial offset $\sqrt{\Delta x^{2}+\Delta y^{2}}$ from the impurity center ($\Delta r=0$). (e) Energy of the impurity bound state for varying tip-sample distance, extracted by fitting a lorentzian curve in 5 intensity plots (see Supplemental Material \cite{Supplement}) similar to (c)-(d).}\label{fig3}
\end{figure*}

It is important to note that the $s_{\pm}$ symmetry of the order parameter in $\mathrm{FeTe}_{0.55}\mathrm{Se}_{0.45}$ can lead to a very similar phenomenology between magnetic and potential scatterers. While in conventional $s$-wave superconductors, magnetic impurities are required to create in-gap (YSR) states, in $s_{\pm}$ superconductors, sub-gap resonances can also occur for non-magnetic scattering centers. This can be shown using different theoretical techniques, including T-matrix method \cite{Bang2009,Kreisel2016}, Bogoliubov-de Gennes equations \cite{Kariyado2010,Tsai2009} and Green's functions \cite{Gastiasoro2013,Ng2009,Beaird2012} applied to multiband systems with $s_{\pm}$ symmetry. The similarity of magnetic and potential scatterers makes a distinction between these cases more challenging (but possible, with an external magnetic field \cite{Fan2020}). In either case, theory predicts energy-symmetric in-gap states with particle-hole asymmetric intensities.

The X-shaped phenomenology of the in-gap states shown above shares also similarities with bound-states that have been observed in Pb/Co/Si(111) stacks \cite{Menard2017}, where they have been interpreted as topological \cite{Menard2017,Garnier2019}. However, as we will show here, in our experiments the single point of zero bias is just one particular case of a manifold of dispersions that depend on the tip-sample distance.

\subsection{Tuning the energy of the in-gap state with the tip}
Figure \ref{fig3}(a) shows an intensity plot of a series of spectra above $\mathbf{r}_{0}$, the location showing the zero-bias impurity state, with changing tip-sample distance (see inset for a schematic). We normalized each spectrum by the normal state resistance $R_{\mathrm{N}}=V_{\mathrm{set}}/I_{\mathrm{set}}$. In order to reduce the distance, we control the tip in constant current feedback and increase the set-point for the current while keeping the voltage bias constant. In addition, we measure the tip-sample distance relative to the set-point: $V_{\mathrm{set}} = 5$ mV, $I_{\mathrm{set}} = 0.4$ nA. Strikingly, we observe a shift in the energy of the in-gap state with varying the tip-sample distance $\Delta d$. When the tip is brought closer to the sample surface, the sub-gap resonances shift towards the Fermi energy (Fig. \ref{fig3}(b)) where they cross and split again.  We also point out that there is a strong particle-hole asymmetry in the intensity of the in-gap resonances. It can be clearly seen in Fig. \ref{fig3}(b) that the relative intensity between the positive ($\mathrm{p}$) and negative ($\mathrm{n}$) resonances ($I_{\mathrm{p}}-I_{\mathrm{n}}$) changes sign after the cross at the Fermi level. To obtain a more complete picture of the tuning of the in-gap states as we vary $\Delta d$, we measured five $dI/dV(\mathbf{r},V)$ maps (each at different tip-sample distance) and analyzed azimuthally-averaged radial profiles through the impurity center (Figs. \ref{fig3}(c)-(d) show two of these profiles. See Supplemental Material \cite{Supplement} for the other 3). We extract the energy of the resonances by Lorentzian fits (Fig. \ref{fig3}(e)), to observe that they cross the Fermi level at the impurity center when being close to the sample. This is the first time that such a crossing has been observed in an unconventional superconductor.

\subsection{Microscopic origin}
The important question that arises is: what tunes the impurity resonances that we observe? In previous experiments with magnetic ad-atoms or ad-molecules on conventional superconductors \cite{Farinacci2018,Malavolti2018}, it has been shown that the force of the tip changes the coupling between moment and substrate, and that the coupling $J$ and the YSR energy could be tuned in this way. In this case, when the energy crosses the Fermi level at the critical coupling $J_{\mathrm{C}}$, a first-order quantum phase transition between the singlet (screened) and the doublet (unscreened) ground state is expected  \cite{Balatsky2006,Farinacci2018}. Very recently, a similar force-based scenario has been reported in different systems involving magnetic ad-atoms on top of superconductors \cite{Huang2019}, including Fe(Te,Se) \cite{Fan2020}. As discussed in the Supplemental Material \cite{Supplement}, a similar scenario can in principle explain the sub-gap dispersion discovered here. However, as the impurity is not loosely bound on top of the surface in the present case, a movement between a sub-surface impurity and the superconductor due to the tip force as the cause for the tuning, seems unlikely. Therefore, we pursue alternative mechanisms. Motivated by the phenomenology of semiconductors \cite{Wijnheijmer2011} or Mott insulators \cite{Battisti2017}, where the tip can act as a local gate electrode (mediated by the poor screening), we propose a similar gating scenario for YSR states in the present case: the electric field of the tip can tune the energy of the impurity state and thus lead to a dispersing YSR state.

First, we note that there can be a significant difference between the work functions of the tip and the sample. Typical work functions are in the range of a few electronvolts, and differences between chemically different materials of the order of an electronvolt are common (see Supplemental Material \cite{Supplement}). Hence, it is possible to have a voltage drop between them that is larger than the applied bias. Secondly, the low carrier density in $\mathrm{FeTe}_{0.55}\mathrm{Se}_{0.45}$ leads to a non-zero screening length giving rise to penetration of the electric field of the tip inside the sample. An estimation of the penetration depth in the sample can be made in the Thomas-Fermi approximation. In this framework, the screening length is given by $\lambda_{\mathrm{TF}}=(\pi a_{0}/4k_{\mathrm{F}})^{1/2}$, where $a_{0}$ is the Bohr radius and $k_{\mathrm{F}}$ the Fermi wave-vector. Using reported parameters \cite{Zhang2018,Wang2018}, this yields  $\lambda_{\mathrm{TF}}=0.5$ nm, which is comparable to the inter-layer distance \cite{Sales2009}. An impurity residing between the topmost layers is thus affected by the electric field of the tip. 

Based on these considerations, we conclude that  it is possible that the tip acts as a local gate electrode that influences the energy levels of the impurity, which in turn influences the energy of the in-gap states, as we will demonstrate in the modelling carried out below. By adjusting the tip-sample distance the field penetration is modulated leading to an energy shift of the in-gap resonances. The spatial dependence can be explained similarly: when moving the tip over the impurity location, we change the local electric field, which is at a maximum when the tip is right above the impurity, with details depending on the tip shape.

\begin{figure*}[ht]
\includegraphics[scale=1]{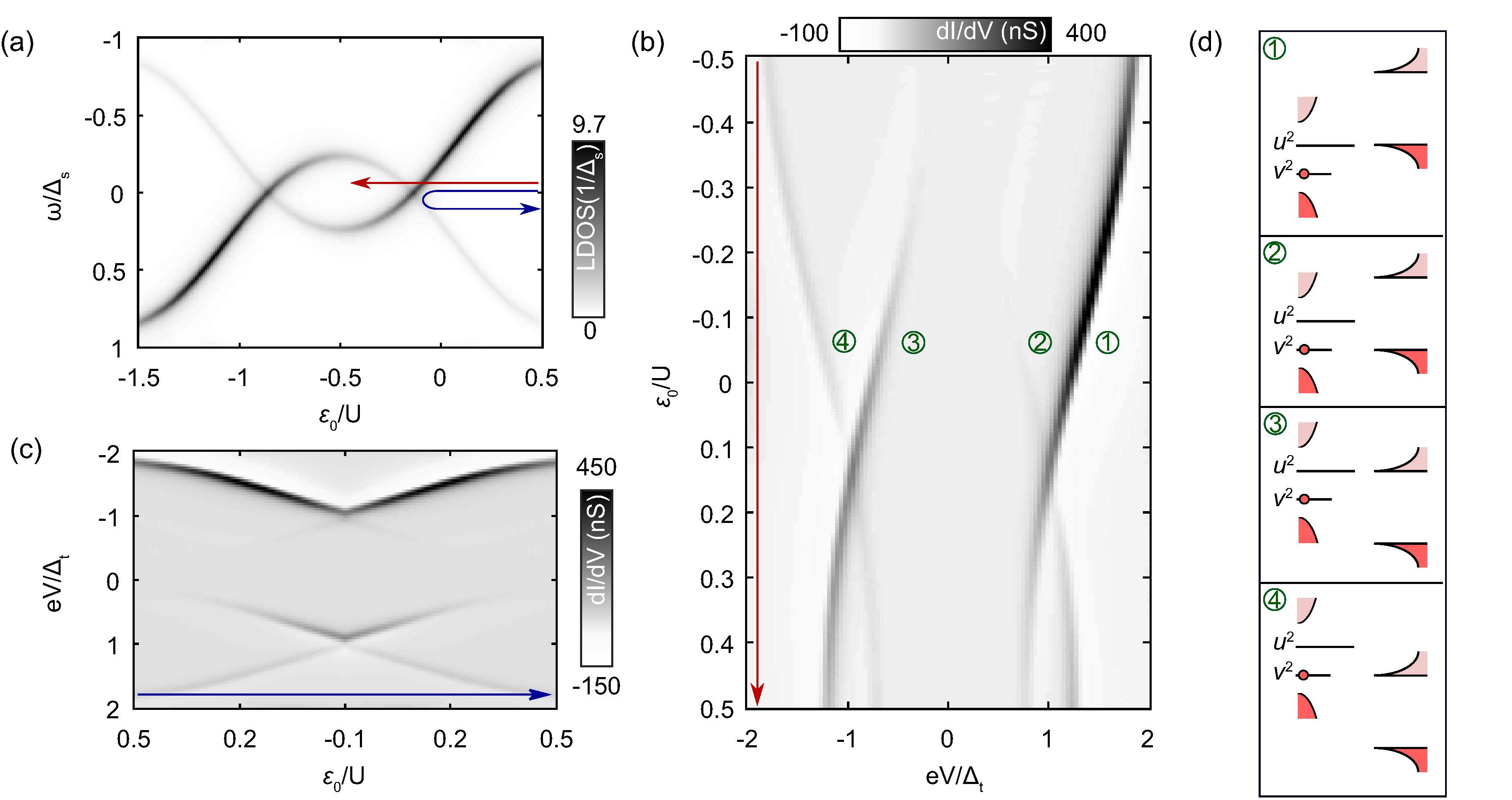}
\caption{Anderson impurity model for Yu-Shiba-Rusinov bound states. (a) Local density of states as a function of level energy $\epsilon_{0}$. The impurity spectral function was calculated within the zero-bandwidth approximation using the Källen-Lehmann spectral representation for the retarded Green’s function (see Supplemental Material \cite{Supplement}). Red and blue lines indicate the two different $\epsilon_{0}$ sweeps plotted in (b) and (c), respectively. (b)-(c) Relaxation dominated tunneling conductance calculated to leading order in the tip-impurity tunneling rate $\Gamma_{\mathrm{t}}$. (b) and (c) are plotted on the same color scale and labels in (b) refer to processes in (d). (d) Guide to the eye for different conductance contributions in (b) and (c). Processes 2-3 require a finite population of the excited state, in this case supplied by temperature. For all panels we use $U = 3$, $\Gamma_{\mathrm{s}} = 1.5$, $\Gamma_{\mathrm{t}} = 0.03$, $\gamma = \Gamma_{\mathrm{r}} = 0.035$ and $k_{\mathrm{B}}T = 0.2$, all in units of $\Delta_{\mathrm{s}}$.}\label{fig4}
\end{figure*} 

\subsection{Gate-tunable single impurity Anderson model}
We model the sub-gap state as a YSR state arising from the magnetic moment of a sub-surface impurity level, whose energy is effectively gated by the tip-induced electric field. It should be noted that the sub-gap states arising in an $s_{\pm}$-wave superconductor from a simple non-magnetic impurity can produce a dispersive cross in the in-gap energies as a function of the impurity potential. However, this is only true for a particular range of potentials, and will not generally trace out a single dispersive cross as a function of the impurity strength \cite{Kariyado2010,Gastiasoro2013}. Therefore, we are led to conclude that the impurity at hand involves a finite magnetic moment. Local impurity-induced magnetic moments may indeed be particularly prominent in correlated systems like FeSe where even nonmagnetic disorder, in conjunction with electron interactions, can generate local moments \cite{song2020}. Because of the magnetic nature of the disorder site, the results of our calculations are qualitatively independent on whether we treat the system as an $s$ or $s_{\pm}$-wave superconductor. For simplicity, we perform our calculations assuming standard $s$-wave pairing.

The superconducting single impurity Anderson model \cite{Anderson1966} involves an impurity level $\epsilon_{0}$ with charging energy $U$ coupled via a tunneling rate $\Gamma_{\mathrm{s}}$  to a superconducting bath with energy gap $\Delta_{\mathrm{s}}$  \cite{Jellinggaard2016,Kirsanskas2015,Zitko2015}. We represent the sample by a simple $s$-wave Bardeen-Cooper-Schrieffer (BCS) superconductor, and use the zero-bandwidth approximation, including only a single spin-degenerate pair of quasiparticles at energy $\Delta_{\mathrm{s}}$ \cite{Kirsanskas2015,Grove-Rasmussen2018}. We further assume that the gating from the tip changes the impurity level $\epsilon_{0}$ linearly with distance.  We then obtain the YSR states  by calculating the local impurity spectral function, $D_{\mathrm{I}} (\omega,\epsilon_{0})$, as a function of $\epsilon_{0}$ (and thus of gating) using the Lehmann representation (see Supplemental Material \cite{Supplement} for details). The result is plotted in Fig. \ref{fig4}(a), where the observed crossing of the sub-gap states  indicates a change between singlet, and a doublet ground state \cite{Bauer2007}. From the spectral function we can determine the current using leading-order perturbation theory in the tunnel coupling connecting the impurity to the tip, $t_{\mathrm{t}}$: 
\begin{multline}
I(V)=\frac{e|t_{\mathrm{t}}|^2}{\hbar} \int D_{\mathrm{t}}(\omega+eV,\Delta_{\mathrm{t}},\gamma_{\mathrm{t}})D_{\mathrm{I}} (\omega,\epsilon_{0})\times \\ [f(\omega,T)-f(\omega+eV,T)]d \omega,
\end{multline} 
here $D_{\mathrm{t}}(\omega,\Delta_{\mathrm{t}},\gamma_{\mathrm{t}})$ is the spectral function of the superconducting tip with a finite quasiparticle broadening incorporated as a Dynes parameter \cite{Dynes1978,Supplement}, $\gamma_{\mathrm{t}}$ and $\hbar$ the reduced Planck's constant. A phenomenological relaxation rate, $\Gamma_{\mathrm{r}}$, is incorporated into the Lehmann representation, (see Supplemental Material \cite{Supplement}), to construct $D_{\mathrm{I}} (\omega,\epsilon_{0})$. This parameter accounts for quasiparticle relaxation of the YSR resonances at $\omega=\pm E_{\mathrm{ig}}$. The validity of the expansion in $\Gamma_{\mathrm{t}}=\pi \nu_{\mathrm{F}} |t_{\mathrm{t}}|^2$, which captures only single electron transport and omits Andreev reflections, rests on the assumption that the sub-gap state thermalizes with rate $\Gamma_{\mathrm{r}}$ between each tunneling event. In the opposite limit, $\Gamma_{\mathrm{t}}\gg\Gamma_{\mathrm{r}}$, transport takes place via Andreev reflections, and the sub-gap conductance peaks at $eV=\pm(\Delta_{\mathrm{t}}+E_{\mathrm{ig}})$ would be of equal magnitude, even though the underlying spectral function is generally asymmetric \cite{Ruby2015}. We note that in principle, shot noise experiments \cite{Bastiaans2018rsi,Bastiaans2018nat,Supplement} or photo-assisted 
tunneling \cite{Peters2020} experiments could differentiate between the two cases. The experimental data shown in Fig. \ref{fig3} display a marked asymmetry, consistent with our assumption of relaxation dominated transport where conductance asymmetry will follow the asymmetry of the underlying spectral function.

Next, we investigate the situations where the tip moves over the impurity along the surface, or towards the impurity as a function of tip-sample  distance $\Delta d$. These situations are marked with blue and red arrows in Fig. \ref{fig4}(a), respectively. In Figs. \ref{fig4}(b)-(c) we then plot sub-gap conductance as a function of level position, $\epsilon_{0}$, corresponding to the red/blue traces, assuming a linear dependence of $\epsilon_{0}$ with tip-sample distance. The agreement between our model and the data is good, both in terms of the energy dispersion and the asymmetry. Also, in both experiment and theory, additional conductance peaks at $eV=\pm(\Delta_{\mathrm{t}}-E_{\mathrm{ig}})$ are visible close to the singlet-doublet phase transition. We interpret these lines as the additional single electron processes shown in Fig. \ref{fig4}(d), which arise from thermal population of the excited state close to the phase transition where $E_{\mathrm{ig}} \lesssim k_{\mathrm{B}} T$. The conductance peaks at $\Delta_{\mathrm{t}} \pm E_{\mathrm{ig}}$ meet at the point where the YSR states cross zero energy, signaling the change between singlet, and doublet ground states, and the asymmetry in intensity between the conductance peaks at $eV=\pm (\Delta_{\mathrm{t}}+E_{\mathrm{ig}})$ switches around. 

The good agreement between this simple model (Figs. \ref{fig4}(b)-(c)) and the data presented in Figs. \ref{fig2}(e) and \ref{fig3}(a), supports our interpretation that the tip exerts an effective gating of the impurity. At this point, we are not able to exclude an alternative scenario, in which the impurity-substrate coupling, $\Gamma_{\mathrm{s}}$, depends monotonically on the tip distance \cite{Farinacci2018}. We discuss alternative scenarios further in the Supplemental Material \cite{Supplement}, but the fact that our impurity is below the surface and the excellent agreement between the model and the data lead us to conclude that the gating scenario is most likely in the present case. 

\section{Conclusion}
In summary, we have reported on the properties of energy symmetric in-gap states in $\mathrm{FeTe}_{0.55}\mathrm{Se}_{0.45}$ that can be tuned through the Fermi level. These states extend over a large ($\sim 10$ nm) area around the center locations. Our data point towards a sub-surface magnetic impurity embedded in a low-density superfluid with large screening length that leads to YSR-like in-gap states. We propose a novel tip-gating mechanism for the dispersion and perform calculations within the single impurity Anderson model that show excellent agreement with the data. Such a mechanism could also play a role in previous experiments on elemental superconductors or heterostructures. How such states are related to the topological superconductivity in $\mathrm{FeTe}_{0.55}\mathrm{Se}_{0.45}$ remains an open question. Our work further shows that one needs to be careful when interpreting zero-bias peaks in putative topological states, and junction resistance dependent experiments are a necessary -- ideally combinded with other techniques such as noise spectroscopy \cite{Jonckheere2019,Golub2011,Bastiaans2019,Supplement}, spin-polarized STM \cite{Wiesendanger2009}, or photon-assisted tunneling \cite{Tang2015} will allow for better understanding. Independent of this, tunable impurity states like the one we report here could offer a platform to study quantum phase transitions, impurity scattering, and the screening behavior of superfluids.

\begin{acknowledgments}
We acknowledge J. de Bruijckere, J. F. Ge, M. H. Fischer, P. Hirschfeld, D. K. Morr, P. Simon, J. Zaanen and H. S. J. van der Zant for fruitful discussions. This work was supported by the European Research Council (ERC StG SpinMelt) and by the Netherlands Organization for Scientific Research (NWO/OCW), as part of the Frontiers of Nanoscience programme, as well as through a Vidi grant (680-47-536). G. D. G. is supported by the Office of Basic Energy Sciences, Materials Sciences and Engineering Division, US Department of Energy (DOE) under contract number de-sc0012704. B. M. A. acknowledges support from the Independent Research Fund Denmark grant number DFF-8021-00047B. The Center for Quantum Devices is funded by the Danish National Research Foundation. D. Cho was supported by the National Research Foundation of Korea (NRF) grant funded by the Korea government (MSIT) (No. 2020R1C1C1007895 and 2017R1A5A1014862) and the Yonsei University Research Fund of 2019-22-0209.
\end{acknowledgments}
%
%

\end{document}


\preprint{}

\title{Supplemental Material:
Spatially dispersing Yu-Shiba-Rusinov states in the unconventional superconductor $\mathrm{FeTe}_{0.55}\mathrm{Se}_{0.45}$}

\author{Damianos Chatzopoulos}
\thanks{These authors contributed equally to this work}
\affiliation{Leiden Institute of Physics, Leiden University, Niels Bohrweg 2, 2333 CA Leiden, The Netherlands}

\author{Doohee Cho}
\thanks{These authors contributed equally to this work}
\affiliation{Leiden Institute of Physics, Leiden University, Niels Bohrweg 2, 2333 CA Leiden, The Netherlands}
\affiliation{Department of Physics, Yonsei University, Seoul 03722, Republic of Korea}

\author{Koen M. Bastiaans}
\thanks{These authors contributed equally to this work}
\affiliation{Leiden Institute of Physics, Leiden University, Niels Bohrweg 2, 2333 CA Leiden, The Netherlands}
\thanks{These authors contributed equally to this work}

\author{Gorm O. Steffensen}
\affiliation{Center for Quantum Devices, Niels Bohr Institute, University of Copenhagen, Universitetsparken 5, 2100 Copenhagen Ø, Denmark.}

\author{Damian Bouwmeester}
\affiliation{Leiden Institute of Physics, Leiden University, Niels Bohrweg 2, 2333 CA Leiden, The Netherlands}
\affiliation{Kavli Institute of Nanoscience, Delft University of Technology, Lorentzweg 1, 2628 CJ Delft, Netherlands}

\author{Alireza Akbari}
\affiliation{Max Planck Institute for the Chemical Physics of Solids, D-01187 Dresden, Germany}
\affiliation{Max Planck POSTECH Center for Complex Phase Materials,and Department of Physics, POSTECH, Pohang, Gyeongbuk 790-784, Korea}

\author{Genda Gu}
\affiliation{Condensed Matter Physics and Materials Science Department, Brookhaven National Laboratory, Upton, NY 11973, USA}

\author{Jens Paaske}
\affiliation{Center for Quantum Devices, Niels Bohr Institute, University of Copenhagen, Universitetsparken 5, 2100 Copenhagen Ø, Denmark.}

\author{Brian M. Andersen}
\affiliation{Center for Quantum Devices, Niels Bohr Institute, University of Copenhagen, Universitetsparken 5, 2100 Copenhagen Ø, Denmark.}

\author{Milan P. Allan}
\email[]{allan@physics.leidenuniv.nl}
\affiliation{Leiden Institute of Physics, Leiden University, Niels Bohrweg 2, 2333 CA Leiden, The Netherlands}

\date{\today}
\maketitle
\renewcommand{\theequation}{S\arabic{equation}}

\section{Deconvolution of spectral density}
In order to obtain the local density of states (LDOS) of the sample we use a deconvolution algorithm which subtracts the density of states of the tip from the measured spectra. We follow the same procedure as in Refs. \cite{Choi2017,Pillet2010}. The first step is to characterize the tip density of states. This is done by taking spectra with a Pb tip on Pb as described elsewhere \cite{Cho2019}. We fit these spectra with 

\begin{equation} 
\left( \frac{dI}{dV} \right)_{\mathrm{Pb-Pb}}=\frac{1}{eR_{\mathrm{N}}} \int \left\{\frac{\partial D_{\mathrm{t}}(\omega+eV)}{\partial V}  \left[ f(\omega)-f(\omega+eV) \right]-D_{\mathrm{t}}(\omega+eV)\frac{\partial f(\omega+eV)}{\partial V} \right\}D_{\mathrm{t}}(\omega)d \omega,
\end{equation}
in order to extract the gap $\Delta_{\mathrm{t}}$ and the broadening term $\gamma_{\mathrm{t}}$ of the tip. For the $D_{\mathrm{t}}(\omega,\Delta_{\mathrm{t}},\gamma_{\mathrm{t}})$, a modified Dynes formula is used
\begin{equation}
D_{\mathrm{t}}(\omega,\Delta_{\mathrm{t}},\gamma_{\mathrm{t}})={\rm Re} \left[ \mathrm{sgn}(\omega)\frac{\omega}{\sqrt{\omega^{2}+2i\gamma_{\mathrm{t}}\omega-\Delta_{\mathrm{t}}^{2}}} \right].
\end{equation}
Good agreement is found for $\Delta_{\mathrm{t}}=1.3$ meV and $\gamma_{\mathrm{t}}=45$ $\mu$eV. Next, we discretize the theoretical tunneling formula for the differential conductance
\begin{equation} 
\left( \frac{dI}{dV} \right)_{j}=\frac{1}{eR_{\mathrm{N}}} \sum_{i} \left\{\frac{\partial D_{\mathrm{t}}(\omega_{i}+eV_{j})}{\partial V_{j}}  \left[ f(\omega_{i})-f(\omega_{i}+eV_{j}) \right]-D_{\mathrm{t}}(\omega_{i}+eV_{j})\frac{\partial f(\omega_{i}+eV{j})}{\partial V{j}} \right\} \delta\omega D_{\mathrm{s}}(\omega_{i}),
\end{equation} 
where $\delta\omega$ is  the energy spacing. Note that we dropped the $\Delta_{\mathrm{t}},\gamma_{\mathrm{t}}$ and $T$ dependence for simplicity. The above formula is then solved in a matrix form in order to obtain $D_{\mathrm{s}}(\omega)$. 

\section{Anderson impurity in a (zero-bandwidth) BCS superconductor}
We model the sub-surface impurity by an Anderson impurity embedded in a simple $s$-wave superconductor. The total Hamiltonian reads: 
\begin{equation}
H_{\mathrm{tot}}=H_{\mathrm{S}}+H_{\mathrm{T}}+H_{\mathrm{IMP}},
\end{equation} 
where $H_{\mathrm{S}}$ is the Hamiltonian for the superconductor, $H_{\mathrm{T}}$ the tunneling contribution and $H_{\mathrm{IMP}}$ describes the impurity. In terms of fermion creation and annihilation operators for the superconductor ($c$) and the impurity level ($d$), each term is given by 
\begin{equation}
H_{\mathrm{S}}=\sum_{k,\sigma}\epsilon_{k}c^{\dagger}_{k\sigma}c_{k\sigma}+\sum_{k}(\Delta_{\mathrm{s}}c^{\dagger}_{k\uparrow}c^{\dagger}_{k\downarrow}+\Delta^\ast_{\mathrm{s}}c_{k\downarrow}c_{k\uparrow}),
\end{equation} 
where $\Delta_{\mathrm{s}}$ is the superconducting gap (here assumed to be real), and 
\begin{equation}
H_{\mathrm{T}}=\sum_{k,\sigma}t_{\mathrm{s}}c^{\dagger}_{k,\sigma}d_{\sigma}+t^{\ast}_{\mathrm{s}}d^{\dagger}_{\sigma}c_{k,\sigma},
\end{equation} 
where we take $t_{\mathrm{s}}=\sqrt{\Gamma_{\mathrm{s}}/\pi \nu_{\mathrm{F}}}$ and $\nu_{\mathrm{F}}$  is the density of states near the Fermi energy for $\Delta_{\mathrm{s}}=0$, which is taken to be constant. 
\begin{equation}
H_{\mathrm{IMP}}=\sum_{\sigma}\epsilon_{0}d^{\dagger}_{\sigma}d_{\sigma}+Ud^{\dagger}_{\uparrow} d_{\uparrow}d^{\dagger}_{\downarrow}d_{\downarrow},
\end{equation} 
with $U$ being the charging energy of the impurity. 

In the normal state, this reduces to the standard Anderson model, including mixed-valence as well as Kondo physics below a characteristic Kondo temperature, $T_{\mathrm{K}}$. In the superconducting state, this simplifies to a single Yu-Shiba-Rusinov (YSR) state, which can be tuned through zero energy by varying $\epsilon_{0}$, as long as $T_{\mathrm{K}}\lesssim 0.3\Delta_{\mathrm{s}}$ \cite{Bauer2007}. Here, we employ the zero-bandwidth approximation (ZBW), which replaces the superconducting by a single pair of BCS quasiparticles at the gap edge, 
\begin{equation}
H^{\mathrm{ZBW}}_{\mathrm{S}}=\Delta_{\mathrm{s}} c^{\dagger}_{\uparrow}c^{\dagger}_{\downarrow}+\Delta_{\mathrm{s}}^{\ast}c_{\downarrow}c_{\uparrow}.
\end{equation}
From comparison with numerical renormalization group calculations, this approximation is known to capture the YSR states very well up to adjustments in the tunnel coupling, the value of which is fitted anyway \cite{Grove-Rasmussen2018}.   

In order to calculate the spectral weights of the impurity resonances in the Anderson model we use the Lehmann spectral representation of the Green’s function. Using the many-body eigenstates $| n \rangle$ and the eigenenergies $E_{n}$ of the ZBW Hamiltonian, the retarded Green’s function on the impurity can be calculated as follows:
\begin{equation}
G^{\mathrm{R}}_{\sigma\sigma^{\prime}}(\omega)=\frac{1}{Z}\sum_{nn^{\prime}}e^{-\beta E_{n}} \left(\frac{\braket{n|d_{\sigma}|n^{\prime}}\braket{n^{\prime}|d^{\dagger}_{\sigma^{\prime}}|n}}{\omega+i\Gamma_{\mathrm{r}}+E_{n}+E_{n^{\prime}}}+\frac{\braket{n|d^{\dagger}_{\sigma^{\prime}}|n^{\prime}}\braket{n^{\prime}|d_{\sigma}|n}}{\omega+i\Gamma_{\mathrm{r}}+E_{n^{\prime}}-E_{n}}\right),
\end{equation}
where $\beta=1/(k_{\mathrm{B}}T)$ and the partition function $Z$ is given by $Z=\sum_{n}e^{-\beta E_{n}}$. Here we have included a phenomenological relaxation rate, $\Gamma_{\mathrm{r}}$, which endows the otherwise sharp bound states by a finite lifetime broadening. The details of this quasiparticle relaxation time are beyond the scope of this work, but our analysis of the tunnelling current assumes it to be larger than or similar to the tip-impurity tunnelling rate, $\Gamma_{\mathrm{t}}$. The local spectral function is expressed in terms of the impurity retarded Green’s function as: 
\begin{equation}
D_{\mathrm{I}}(\omega,\epsilon_{0})=-\frac{1}{\pi}{\rm Im}\left[\sum_{\sigma}G_{\sigma\sigma}^{\mathrm{R}}(\omega)\right].
\end{equation} 
Figs. \ref{figS1} and \ref{figS2} show how this local density of (YSR) states changes with level energy, $\epsilon_{0}$, and tunnelling rate $\Gamma_{\mathrm{s}}$, respectively. The quantum phase transition is revealed as the point by which the YSR state crosses zero energy and the spectral weight is exchanged between positive, and negative energy states. This asymmetry in spectral weight has a simple origin within the ZBW model. The observed excitation is between a doublet state, which is simply a single electron on the dot ($\ket{\uparrow_{\mathrm{d}},0_{\mathrm{qp}}}$, $\ket{\downarrow_{\mathrm{d}},0_{\mathrm{qp}}}$), and a more complicated singlet state. This singlet state continuously evolves from an empty impurity level, $\ket{0_{\mathrm{d}},0_{\mathrm{qp}}}$, at $\epsilon_0/U\gg 0$ to a doubly occupied level, $\ket{2_{\mathrm{d}},0_{\mathrm{qp}}}$, at $\epsilon_0/U \ll -1$. In between, these two singlets are mixed with the YSR singlet, formed by the singly occupied impurity level and the single BCS quasiparticle doublet, i.e.$\ket{\uparrow_{\mathrm{d}},\downarrow_{\mathrm{qp}}}-\ket{\downarrow_{\mathrm{d}},\uparrow_{\mathrm{qp}}}$ states. As the positive (negative) part of the spectral function is related to excitations from adding an electron (hole) to the ground state, this part is largest for $\epsilon_0/U \gg 0$ ($\epsilon_0/U\ll -1$). Precisely at $\epsilon_0/U = -0.5$ the state is equally composed of empty and doubly occupied components and as such the spectrum shows no asymmetry. The larger $\Gamma_{\mathrm{s}}$,  the larger the gate-range  around $\epsilon_0/U=-0.5$ in which the YSR singlet component dominates this excited singlet state, and the lower its energy. This explains both the closing of the doublet sector and the increased spectral symmetry as $\Gamma_{\mathrm{s}}/\Delta_{\mathrm{s}}$ increases, observed in Fig. \ref{figS1}.

As is evident from these two figures, the phase transition may be induced either by tuning the level energy, $\epsilon_0$, or the tunnel coupling, $t_{\mathrm{s}}$. As explained in the main text, previous experiments \cite{Farinacci2018} have reported a tip-induced tuning of $t_{\mathrm{s}}$ by assuming a force acting on the impurity from the tip. Whereas this seems feasible for a flexible molecule placed on the surface, we believe that this possibility is less likely in the present case, where the impurity is sub-surfactant. No indication of lattice deformation when tip-sample distance changes (pushing or pulling) is observed in our experiments. Therefore, we consider the effective gating mechanism much more likely to explain the observed tuning of the sub-gap states. 

\section{Azimuthally-averaged radial profiles and spatial mapping of the YSR resonances}
For completeness in Figs. \ref{figS3}(a)-(e) we present all the five azimuthally-averaged radial profiles, each at different tip-sample distance. In Fig. \ref{figS3}(f) we also show a spatial map of the YSR resonances obtained by fitting a lorentzian. 

\section{Noise spectroscopy}
As an extra verification for the performance of our junction, we used the Scanning Tunneling Noise Microscopy (STNM) technique \cite{Bastiaans2018rsi,Bastiaans2018nat} to measure the current noise as function of energy (noise spectroscopy) on a single location on the $\mathrm{FeTe}_{0.55}\mathrm{Se}_{0.45}$ surface, not at an impurity location, with the aim to look for the doubling of current noise due to Andreev reflections. Fig. \ref{figS4}(a) shows the measured current noise power as function of applied bias $S(V)$, with the out of tunneling noise subtracted to remove the thermal noise component and input noise of the amplifier $S(V)-S(0)$. The dashed lines indicate the predicted shot noise curve \cite{Blanter2000,deJong1994,Cuevas1999} for tunneling of single electron charge ($e$) and double electron charge ($2e$). At a bias voltage larger than the superconducting gap energies, $eV>|\Delta_{\mathrm{t}}+\Delta_{\mathrm{s}}|$, the measured noise power data follows the predicted noise power for single electron tunneling. When the bias is lowered below the superconducting gap energies, $eV<|\Delta_{\mathrm{t}}+\Delta_{\mathrm{s}}|$, the current noise clearly deviates from single electron charge transfer, showing a doubling of noise power to the $2e$ line, consistent with the appearance of Andreev reflection processes \cite{Bastiaans2019,deJong1994,Cuevas1999}. To show the effective charge transferred between the Pb tip and $\mathrm{FeTe}_{0.55}\mathrm{Se}_{0.45}$ surface we divide the measured noise power by the full Poissonian noise for single electron charge transport $S=2e|I|$, which is shown in Fig. \ref{figS4}(b). This illustrates a clear step from $e$ to $2e$ charge transfer when the bias is lowered below the superconducting gap energies $eV<|\Delta_{\mathrm{t}}+\Delta_{\mathrm{s}}|$, demonstrating that the tunneling current is now effectively carried by double charge quanta due to Andreev reflection processes that start to dominate. This is the first time such noise enhancement due to Andreev reflection processes is shown in a junction containing an unconventional superconducting electrode (the $\mathrm{FeTe}_{0.55}\mathrm{Se}_{0.45}$ surface), opening a potential new path for further investigation of the YSR or Majorana states by the means of noise spectroscopy.

\section{Estimation of the potential drop in the vacuum barrier between tip and sample}
In an STM experiment there is a potential drop in the vacuum tunneling barrier \cite{Battisti2017} due to the work function difference between tip ($W_{\mathrm{t}}$) and sample ($W_{\mathrm{s}}$), as illustrated in Fig. \ref{figS5}. As we explain in the main text, this potential drop results into an effective electric field penetration in the sample which acts as a local gate. An estimation of the work function difference yields that in our experiment the tip has a larger work function than the sample. In more detail, we are using a Pt/Ir tip coated with Pb. From literature we find that the work functions are $W_{\mathrm{Pt/Ir}}=5-6$ eV and $W_{\mathrm{Pb}}=4.25$ eV, respectively. This yields a rough estimation of the work function of the tip $W_{\mathrm{tip}}=4.5-5$ eV. Concerning the top surface of our sample it consists of Se, $W_{\mathrm{Se}}=5.9$ eV and Te, $W_{\mathrm{Te}}=4.95$ eV atoms. However, previous STM experiments \cite{Massee2011} show that Fe(Se,Te) has a significantly smaller work function of $\sim 3$ eV. This yields a work function difference of the order of 1 eV.


%

\renewcommand{\thefigure}{S\arabic{figure}} 

\begin{figure}[ht]
\includegraphics[scale=1]{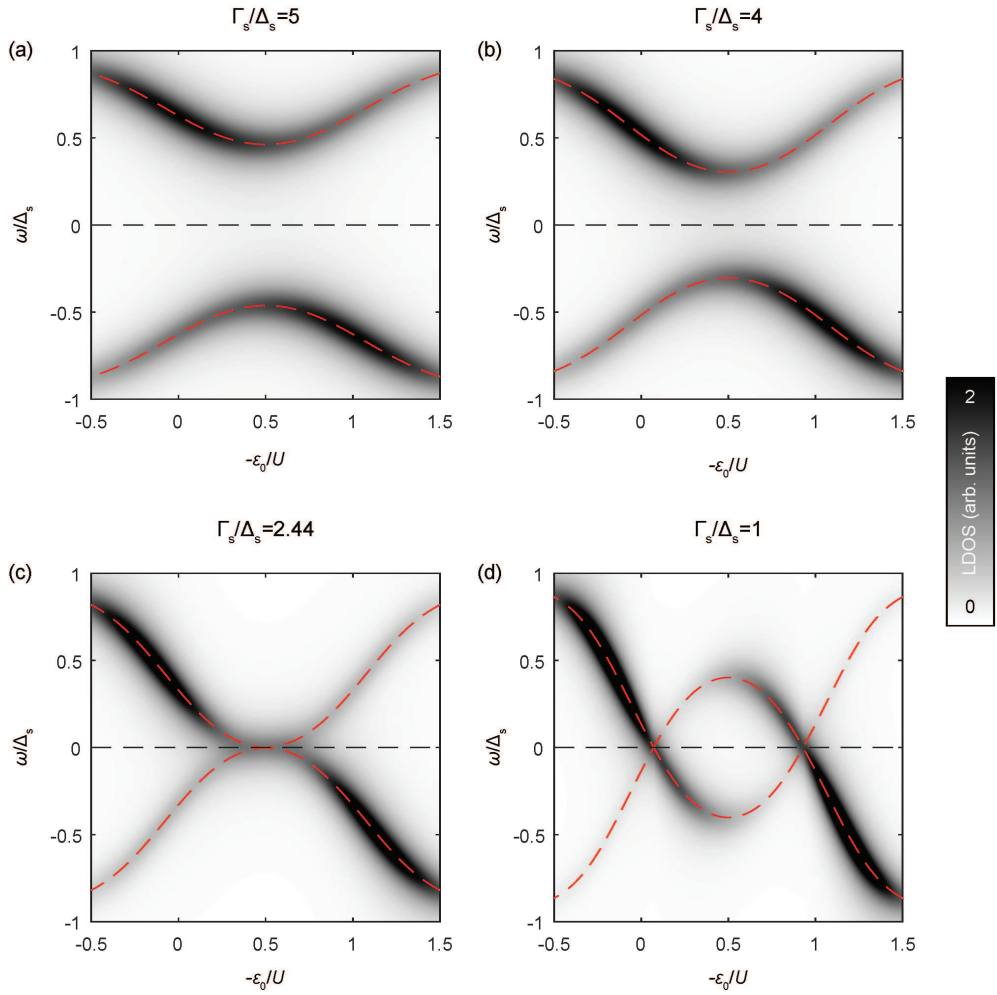}
\caption{Local density of (sub-gap) states as a function of energy ($\omega$) and impurity level energy ($\epsilon_{0}$) for different values of the tunneling rate $\Gamma_{\mathrm{s}}$ as indicated on the top of each panel. The (YSR) bound state energy is highlighted with red dashed lines. In all panels we used $U/\Gamma_{\mathrm{s}}=3$, together with a phenomenological quasiparticle relaxation rate, $\Gamma_{\mathrm{r}}=0.1\Delta_{\mathrm{s}}$.}\label{figS1}
\end{figure}

\begin{figure}[ht]
\includegraphics[scale=1]{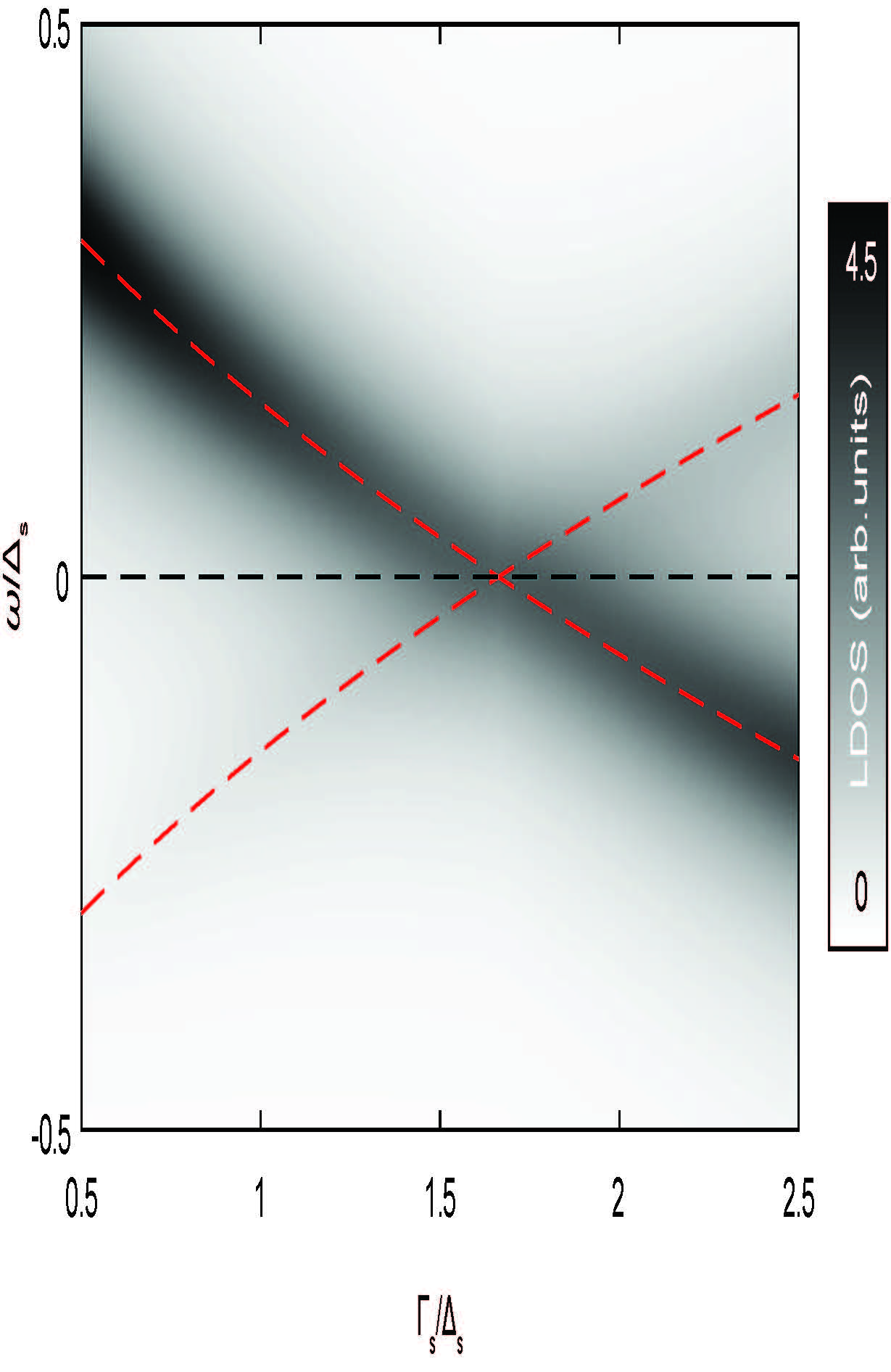}
\caption{
Local density of (sub-gap) states as a function of energy ($\omega$) and tunneling rate ($\Gamma_{\mathrm{s}}$). The (YSR) bound state energy is highlighted with red dashed lines. For the simulations we used $U/\Delta_{\mathrm{s}}=3$, $\epsilon_{0}/\Delta_{\mathrm{s}}=-2.5$, together with a phenomenological quasiparticle relaxation rate $\Gamma_{\mathrm{r}}=0.1\Delta_{\mathrm{s}}$.}\label{figS2}
\end{figure}

\begin{figure}[ht]
\includegraphics[scale=1]{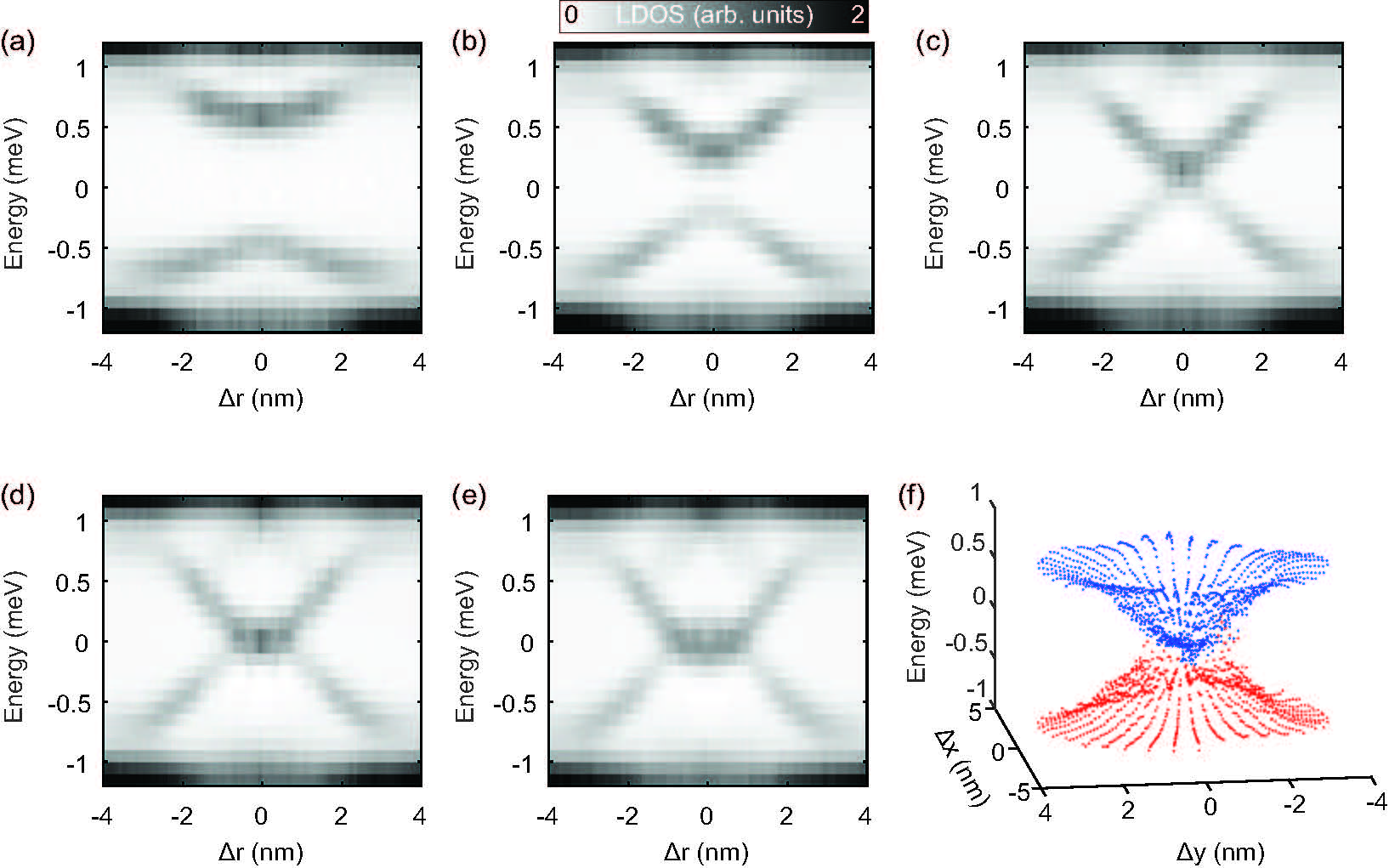}
\caption{Azimuthally-averaged radial profiles and spatial mapping of the YSR resonances.(a)-(e) Azimuthally-averaged radial cuts for $\Delta d(\mathrm{\AA})=1.1,-0.2,-0.7,-1.2$ and $-1.4$, respectively. Note that the colorbar applies to all panels. (g) Spatial mapping of the YSR energy resonances. The energy of each resonance is obtained by fitting a lorentzian at each point of a $\mathrm{LDOS}(\mathbf{r},\omega)$ at $\Delta d(\mathrm{\AA})=-1.2$.}\label{figS3}
\end{figure}

\begin{figure}[ht]
\includegraphics[scale=1]{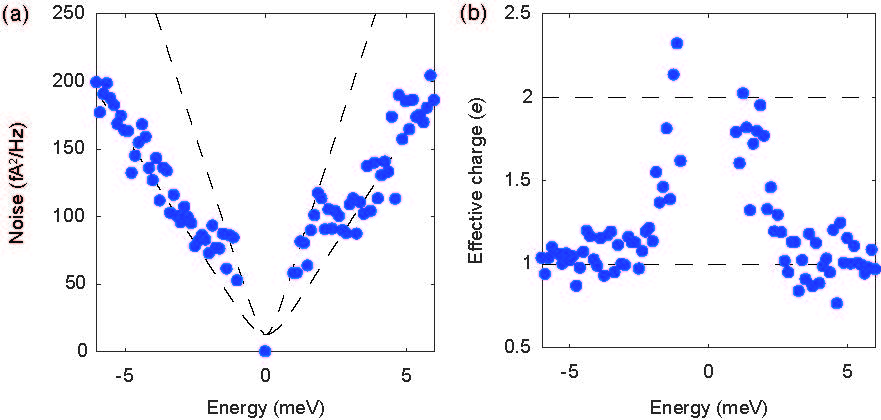}
\caption{Shot-noise measurement on $\mathrm{FeTe}_{0.55}\mathrm{Se}_{0.45}$. (a) Measured current noise power (blue dots) as function of applied bias on the $\mathrm{FeTe}_{0.55}\mathrm{Se}_{0.45}$ surface, while keeping a constant junction resistance of $R_{\mathrm{N}} = 10$ M$\Omega$. The black dashed lines represent the expected current noise power for $q=e$ and $q=2e$ tunneling. (b) Effective charge transferred between the Pb tip and $\mathrm{FeTe}_{0.55}\mathrm{Se}_{0.45}$ surface tunnel junction, obtained by dividing the measured current noise power by the full poissonian noise $2e|I|$, similar to the Fano factor. Dashed lines indicate $q=e$ and $q=2e$.}\label{figS4}
\end{figure}

\begin{figure}[ht]
\includegraphics[scale=1]{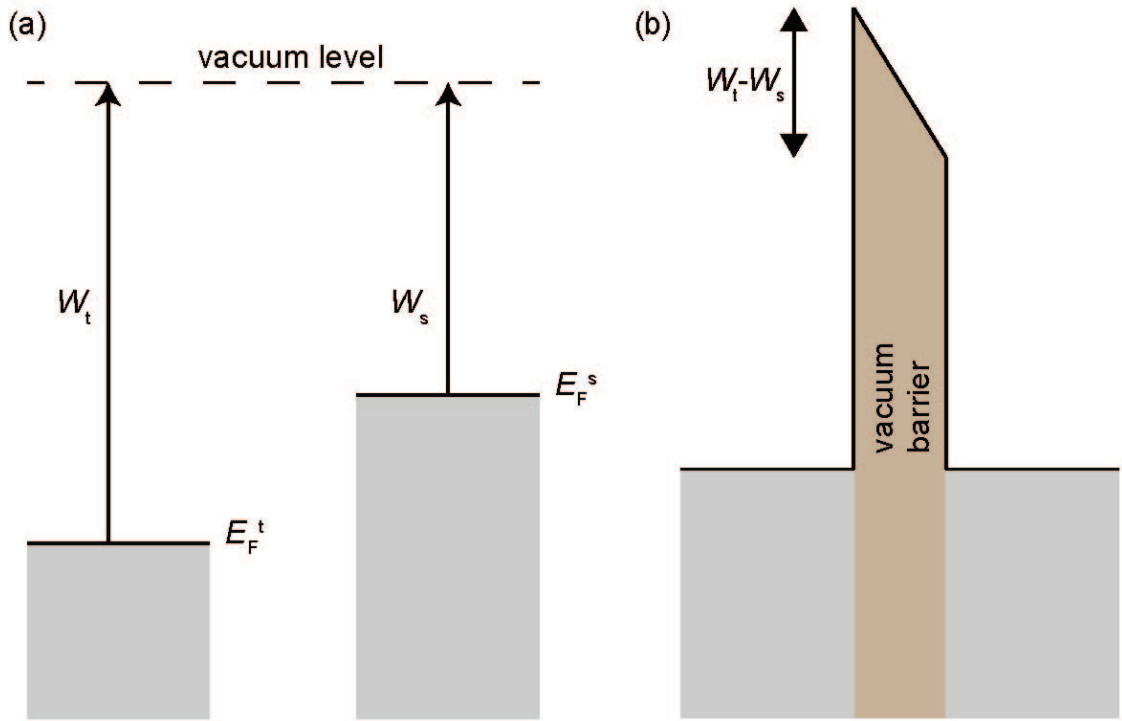}
\caption{Potential drop in the vacuum barrier in a STM junction. (a) Schematic representation of the tip and sample density of states when they are isolated. Grey shaded areas indicate filled states whereas the work function is represented with arrows. (b) When a tunneling contact is formed due to the work function difference there is a potential drop in the vacuum barrier.}\label{figS5}
\end{figure}